\documentstyle[preprint,aps]{revtex}
\begin{document}
\title{Effective one-dimensionality of AC hopping conduction in
the extreme disorder limit}
\draft
\author{Jeppe C. Dyre and Thomas B. Schr{\o}der}
\address{Department of Mathematics and Physics (IMFUFA), 
Roskilde University, POBox 260, DK-4000 Roskilde, DENMARK,
E-mail: dyre@mmf.ruc.dk}
\date{December 7, 1995}
\maketitle{}
\begin{abstract}
It is argued that in the limit of extreme disorder AC
hopping is dominated by "percolation paths".  Modelling a
percolation path as a one-dimensional path with a sharp jump
rate cut-off leads to an expression for the universal AC
conductivity, that fits 
computer simulations in two and three dimensions better than the
effective medium approximation.
\end{abstract}
\pacs{72.20.-i; 72.80.Ng; 66.30.Dn}
While ordered solids show no frequency-dependence of 
their conductivity at frequencies below phonon frequencies,
disordered solids are characterized by AC conductivity 
that varies as an approximate
power-law of frequency 
\cite{jon,owe77,ing87,ell90,ang90,fun93}.  
The exponent is usually 
less than one, but
often quite close to one.  As the frequency goes to zero
 the conductivity stabilizes and 
becomes frequency-independent.  The characteristic frequency
marking
the onset of DC conduction has roughly the same 
temperature-dependence as the DC conductivity.  
These features are observed
universally for electronically conducting disordered 
solids like amorphous
semiconductors \cite{jon,owe77,ell90,lon82}, 
polymers \cite{bly79,kre90}, doped crystalline
semiconductors
at helium temperatures \cite{pol61} (where the random positions
of the dopant atoms becomes
important), or high temperature superconductors above $T_c$
\cite{mol93},
as well as for ionically conducting disordered solids like 
glasses or polymers \cite{owe77,ing87,ang90,fun93}.  

The standard model for AC conduction in disordered solids 
is the hopping model
\cite{ale81,bot85,hau87,nik87,dyr88,bou90,hug95}.
The simplest version is hopping of
non-interacting charge carriers on a regular lattice with random
symmetric nearest neighbor jump rates; this is the model to
be
studied below.  Alternatively, a macroscopic approach may be
adopted by
considering Maxwell's equations for a solid with spatially
randomly
varying conductivity \cite{fis86,dyr93a,dyr93b}.  For both models
the limit of 
extreme disorder may be studied by letting the temperature go to
zero
when the jump rates/the macroscopic
conductivities are thermally activated with randomly varying
activation energies.
It has recently been shown 
by computer simulations \cite{dyr93a,dyr93b,dyr94}
that in the low temperature limit, 
the AC conductivity becomes universal in both models, 
i.e., independent of the activation energy probability
distribution, $p(E)$. The effective medium approximation (EMA) 
for both models predicts the same
universal AC conductivity in the extreme disorder limit (in more
than one dimension).
If ${\tilde \sigma}$ denotes the conductivity relative to the 
DC conductivity and ${\tilde s}$ is a suitably
scaled dimensionless imaginary frequency (``Laplace frequency''),
the EMA universality equation \cite{dyr93a,dyr93b,dyr94}, first
derived by Bryksin for the model of
electrons tunnelling between randomly localized positions 
\cite{bot85,bry80a}, is 

\begin{equation}\label{1}
{\tilde \sigma} \ \ln {\tilde \sigma}\ =\ {\tilde s}\,.
\end{equation}
Computer simulations of the macroscopic model
\cite{dyr93a,dyr93b}
showed that this equation works very well in
two and three dimensions.  
The situation is different for hopping.  
While the existence of universality was confirmed
in two dimensions, it was found \cite{dyr94}
that the onset of AC conduction is considerably smoother than
predicted by Eq.\ (\ref{1}).  No simulations of AC hopping
in three dimensions have so far
been reported studying the extreme disorder limit.

For hopping, the EMA is thus {\bf qualitatively}
correct by predicting universality in the extreme disorder limit,
but {\bf quantitatively} inaccurate.  This is perhaps not
surprising, since universality at extreme disorder
is due to the dominance of percolation effects \cite{dyr94} and
the EMA is a mean-field theory for percolation (which is a
critical phenomenon).  The
EMA replaces the disordered solid by an ``effective'' homogeneous
solid with characteristics
determined by a self-consistency condition.  Such an ordered
medium cannot {\it a priori} be expected 
to accurately represent conduction along
the optimal "percolation" paths of an extremely disordered medium
\cite{hun95}.  

An alternative to the EMA is the ``percolation path
approximation''(PPA) proposed for the macroscopic model
\cite{dyr93b}.
According to the PPA, in the extreme disorder limit the AC
conductivity is equal to that of
a one-dimensional model with a sharp activation energy 
cut-off; this explains the existence of universality
\cite{dyr93b}.  The idea
is that percolation paths, which at extreme
disorder dominate conduction in more than one dimension, 
have two characteristics:
They are essentially one-dimensional and they only involve 
activation energies up to the ``percolation energy'', 
$E_c$, defined from the bond percolation threshold, $p_c$, by

\begin{equation}\label{1a}
p_c\ =\ \int_{-\infty}^{E_c}p(E)dE\,.
\end{equation}
The percolation energy is the activation energy of the
DC conductivity \cite{amb71,shk71}.
For the macroscopic model the PPA leads to 
${\tilde \sigma}={\tilde s}/\ln(1+{\tilde s})$, which is close to
Eq.\ (\ref{1}) and gives a
good fit to simulations in more than one dimension
\cite{dyr93b}.  

The hopping version of PPA is not analytically
solvable.  Below we 
derive an approximation to the hopping PPA utilizing the 
{\bf one-dimensional} EMA, which is known to work well
\cite{bry80b}.
Despite the above objections against the EMA,
this procedure does make sense, because in one dimension
the ``effective'' homogeneous medium 
still is one-dimensional, of course. 
We show by computer simulations that the one-dimensional EMA, 
henceforth referred identified with the PPA, gives a better
representation of low-temperature
AC hopping in two and three dimensions than Eq.\ (\ref{1}). 
Finally, we briefly discuss the consequences of these findings.

To arrive at the PPA, hopping in one dimension 
with a sharp energy barrier cut-off,
$p(E)=0$ for
$E>E_c$ while $p(E_c) > 0$, is addressed.  
In the ``rationalized''
unit system where the conductivity for
a homogeneous system is equal to the jump rate \cite{hau87}, 
the EMA equation for the AC conductivity $\sigma(s)$
in one dimension 
\cite{hug95,dyr94,bry80b,sum81,web81,mov81,oda81,sah83} is
[where $s=i\omega$ is the Laplace frequency, 
$\Gamma$ is the jump rate, and the brackets denote an
average over the jump rate probability distribution]

\begin{equation}\label{2}
\left\langle 
\frac{\Gamma - \sigma}{\sigma + (1-s {\tilde G})(\Gamma -
\sigma)} 
\right\rangle\ =\ 0\,.
\end{equation}
The quantity $s {\tilde G}$ is $s$ times the diagonal element
of the Green's function for a random walk on a one-dimensional
lattice
with uniform jump rate $\sigma$
[the ``effective medium'']; $s {\tilde G}$ is given 
\cite{dyr94,bry80b,sah83} by

\begin{equation}\label{3}
s {\tilde G}\ =\ \left( 1+  \frac{4 \sigma}{s}
 \right)^{-1/2}\,.
\end{equation}
We are only concerned here with relatively low frequencies
where $s{\tilde G}<<1$.
To lowest order in $s{\tilde G}$ Eq.\ (\ref{2}) may be rewritten
\cite{dyr94} 

\begin{equation}\label{4}
\frac{1}{\sigma}\ =\left\langle 
\frac{1}{\Gamma+s{\tilde G} \sigma}
\right\rangle\,.
\end{equation}
The right hand side may be expanded as a power series in 
$s{\tilde G}\sigma$, leading to

\begin{equation}\label{5}
\frac{1}{\sigma}\ =\sum_{n=0}^{\infty}{(-s{\tilde G} \sigma)}^n
\left\langle \Gamma^{-(n+1)}\right\rangle\,.
\end{equation}
Since $\Gamma=\Gamma_0\exp(-\beta E)$, where $\beta$ 
is the inverse temperature, the average
$\left\langle \Gamma^{-(n+1)}\right\rangle$ is easily evaluated
in the low temperature limit:
If $\tilde\beta=\beta/p(E_c)$, one finds to leading order in
$1/{\tilde\beta}$ 
$\left\langle \Gamma^{-(n+1)}\right\rangle=
\Gamma(E_c)^{-(n+1)}/[(n+1){\tilde\beta}]$.
When this is substituted into Eq.\ (\ref{5}) the following
equation is obtained

\begin{equation}\label{7}
\frac{1}{\sigma}\ =\ \frac{1}{{\tilde\beta}}
\frac{1}{s{\tilde G} \sigma}\ 
\ln \left[ 1+\frac{s{\tilde G} \sigma}{\Gamma(E_c)}\right]\,.
\end{equation}
Letting $s$ go to zero we find
$\sigma(0)={\tilde\beta}\ \Gamma(E_c)$.   Introducing
the dimensionless Laplace frequency, 

\begin{equation}\label{7a}
{\tilde s}=\frac{{{\tilde\beta}}^2}{4\ \sigma(0)}\ s\,, 
\end{equation}
whenever $s\tilde G<<1$ Eq.\ (\ref{3}) implies
${\tilde\beta}s{\tilde G}=\sqrt{{\tilde s}/{\tilde\sigma}}$,
where ${\tilde \sigma}=\sigma/\sigma(0)$.  
Substituting this
and $\Gamma(E_c)=\sigma(0)/{\tilde\beta}$
into Eq.\ (\ref{7}) 
finally leads to the PPA expression for hopping,

\begin{equation}\label{8}
\sqrt{{\tilde \sigma}}\ \ln \left[ 
1+\sqrt{{\tilde s}{\tilde \sigma}}
\right]\ =\ \sqrt{{\tilde s}}\,.
\end{equation}
Due to the factor ${\tilde\beta}^2$ in Eq.\
(\ref{7a}), as the temperature is lowered towards zero the
condition 
$s{\tilde G}<<1$ is obeyed in a wider and wider range of
frequencies around
the onset of AC conduction.

We have carried out computer simulations of low-temperature AC
hopping in one,
two and three dimensions using the Fogelholm algorithm
\cite{fog80}
to reduce the AC Miller-Abrahams electrical equivalent circuit of
hopping \cite{mil60,pol73,sum82} according to a recently
proposed scheme \cite{dyr94}.
To speed up the calculations the lowest jump rates were set to
zero; it was carefully checked that this does not affect the
conductivity.
Figure one shows the results of computer simulations of 
low-temperature AC hopping in three dimensions.
Results are shown for averages of 100 simulations of the AC
conductivity at real Laplace frequencies for four different
activation energy probability distributions at the following
``reduced'' inverse temperatures:
(a) ${\tilde \beta}=80$;
(b) ${\tilde \beta}=160$; and
(c) ${\tilde \beta}=320$.
The full curve is the PPA (Eq.\ (\ref{8})) while the dashed curve
is the EMA (Eq.\ (\ref{1})).
Empirical rescalings of the data were allowed in order to focus
only on the {\bf shape} of the conductivity curves.
Figure one shows that universality is approached as the
temperature goes to zero and that the PPA gives a good fit to the
universal AC conductivity in three dimensions.

The universality may be studied without use of empirical
rescalings by plotting the slope of the Log-Log plot,
${\rm d\ Log}{\tilde\sigma}/{\rm d\ Log}{\tilde s}$, as function
of $\tilde\sigma$.
This is done in Fig. 2 for data from computer
simulations in one, two and three dimensions.
The computer simulations in one dimension were carried out for 
systems with a sharp activation energy cut-off, to
check the validity of Eq.\ (\ref{8}).
The full curve is the PPA prediction and the dashed curve is the
EMA prediction in more than one dimension (Eq.\ (\ref{1})).
Clearly, the PPA works better than the EMA in two and
three dimensions; in fact, the PPA works very well in three
dimensions.

As mentioned in the introduction, for the macroscopic model the
EMA works very well in two and three dimensions [it is exact
in one dimension].  For the macroscopic
model the EMA universality prediction (Eq.\ (\ref{1})) is very
close to the macroscopic PPA \cite{dyr93b}.  In view of the above
presented results for 
hopping, it now appears that the EMA works well for the
macroscopic 
model because the EMA {\bf happens to be} close to the PPA,
in contrast to what is the case for hopping.  The two models
have the common characteristic that, in the extreme disorder
limit, conduction becomes essentially
one-dimensional and that consequently the PPA gives a good
description of the universal low-temperature AC conductivity for both
models.

To summarize, we have presented evidence showing that the PPA,
despite being based on a naive one-dimensional picture of
percolation,
is a good model for AC hopping conduction in 
the extreme disorder limit in two and three dimensions. 
These findings have important experimental consequences.
The low-temperature universal conductivity
of the macroscopic model is {\bf different}
from that of the hopping model.  The macroscopic model
incorporates Coulomb interactions via Maxwell's
equations, while hopping traditionally is concerned with
non-interacting particles. 
Consequently, it is in principle possible to determine the
relevance
of Coulomb interactions by measuring the low-temperature
AC conductivity.

\acknowledgements
 This work was supported by the Danish Natural Science Research
Council.

\begin{figure}
\caption[ligegyldigt]

Log-Log plot (base 10) of computer simulations (symbols) of
low-temperature AC conductivity in three dimensions
at real Laplace
frequencies for four different activation energy probability
distributions, compared to the PPA 
(Eq.\ (\ref{8}), full curve) and the EMA (Eq.\ (\ref{1}), dashed
curve).
The dimensionless ``reduced'' Laplace frequency, ${\tilde s}$,
is defined by Eq.\ (\ref{7a}) - however, an
empirical rescaling was allowed to focus exclusively on
the {\bf shape} of the conductivity curve;
${\tilde \sigma}$ is defined
by ${\tilde \sigma}=\sigma(s)/\sigma(0)$.
The jump rates are
$\Gamma=\Gamma_0 \exp(-\beta E)$, where
the activation energy $E$ is chosen
randomly according to the following probability distributions
\cite{dyr93b}:
Asymmetric Gaussian 
[$p(E) \propto  \exp(-E^2/2),\ 0<E<\infty$] 
($\times$);
Cauchy 
[$p(E) \propto  1/(1+E^2),\ 0<E<\infty$]
($+$);
Exponential
[$p(E) \propto  \exp(-E),\ 0<E<\infty$]
($\Diamond$);
and Box
[$p(E)=1,\ 0<E<1$]
($\Box$).
To speed up the calculations all jump rates with activation
energy larger than $E_c+6.4/\beta$ were set to zero [$E_c$ is
defined from the bond percolation threshold in 
Eq.\ (\ref{1a})].
In terms of the dimensionless inverse temperature
${\tilde\beta=\beta/p(E_c)}$ the figure shows data for 
100 averages of simulations of cubic
lattices with sidelength $N$ where
(a) ${\tilde\beta}=80\ [N=29]$;
(b) ${\tilde\beta}=160\ [N=54]$;
(c) ${\tilde\beta}=320\ [N=100]$.
\end{figure}
\begin{figure}
\caption[ligegyldigt]

The slope of the Log-Log plot of ${\tilde\sigma}({\tilde s})$ at
real Laplace frequencies,
${\rm d\ Log_{10}}{\tilde\sigma}/{\rm d\ Log_{10}}{\tilde s}$,
as function of ${\rm Log_{10}}{\tilde\sigma}$ for simulations
(symbols, as in Fig. 1) at the inverse dimensionless temperature
$\tilde\beta=320$ in: (a) one dimension
[100 averages of $8192$ lattices]; (b)
two dimensions [10 averages of $880X880$ lattices]; (c) three
dimensions [100 averages of $100X100X100$ lattices] for the four
activation energy probability distributions of Fig. 1.
The simulations in one dimension were carried out with a sharp
activation energy cut-off at $E=1$ in order to show the validity
of Eq.\ (\ref{8}) in one dimension.
In two and three dimensions, to speed up the calculations, all
jump rates with activation energy larger than $E_c+6.4/\beta$
were set to zero.
The simulations are compared to the predictions
of the PPA (full curve, Eq.\ (\ref{8})) and the EMA (dashed
curve, Eq.\ (\ref{1})).
Both the EMA and the PPA predicts that the slope of the Log-Log
plot goes to one as ${\tilde s}\rightarrow\infty$, but
the PPA works better than the EMA in two dimensions and much
better than the EMA in three dimensions.
\end{figure}

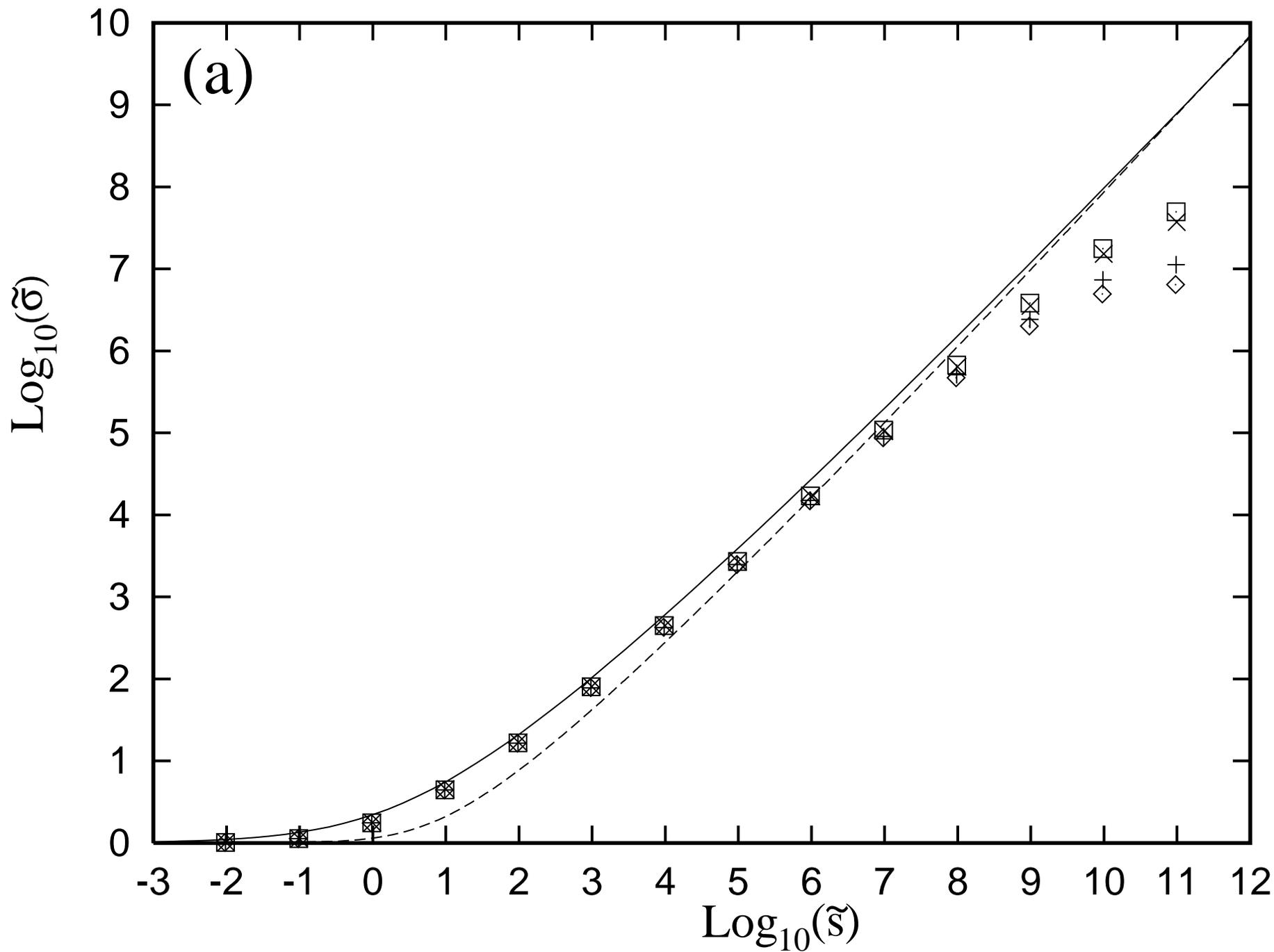
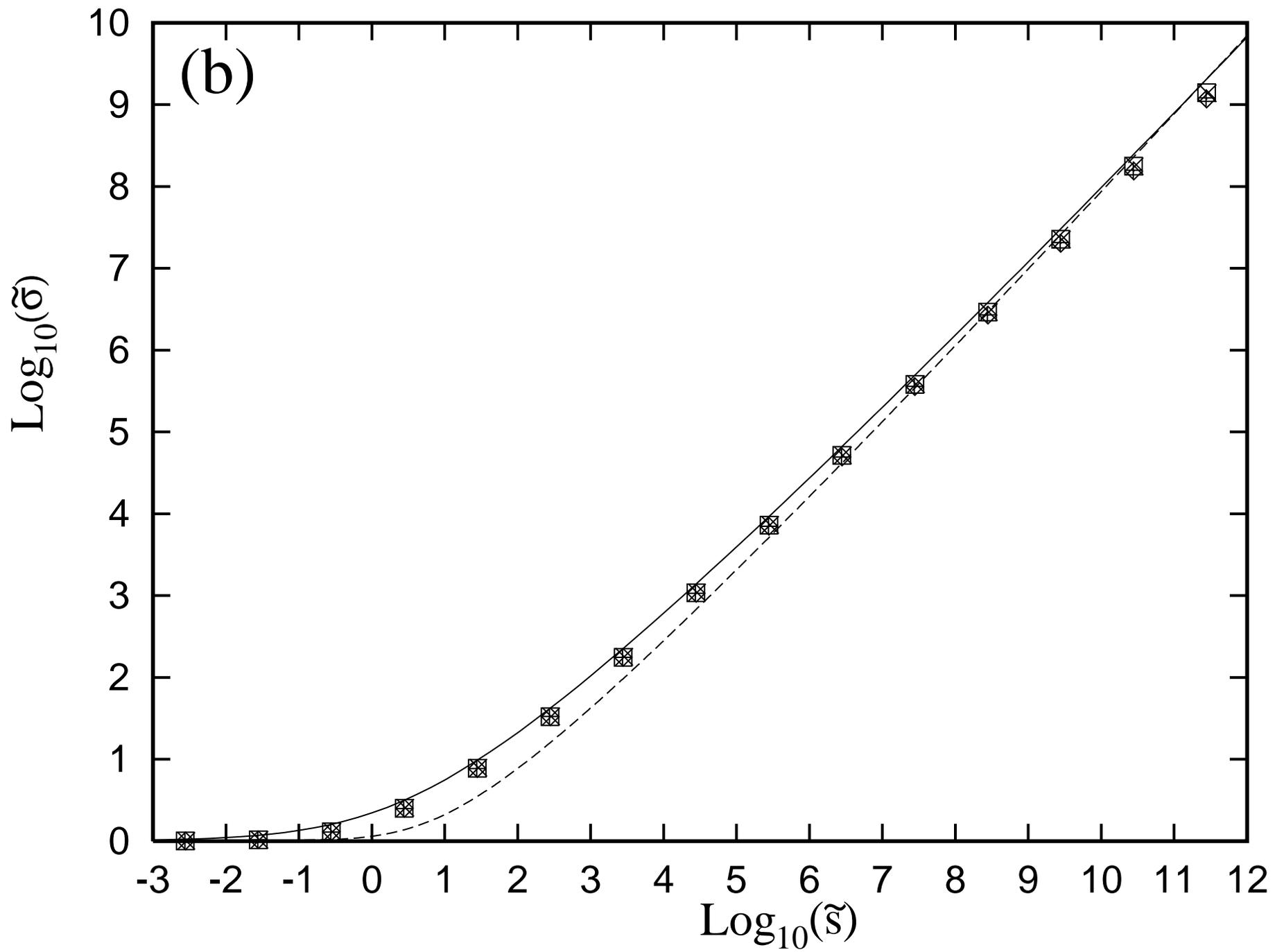
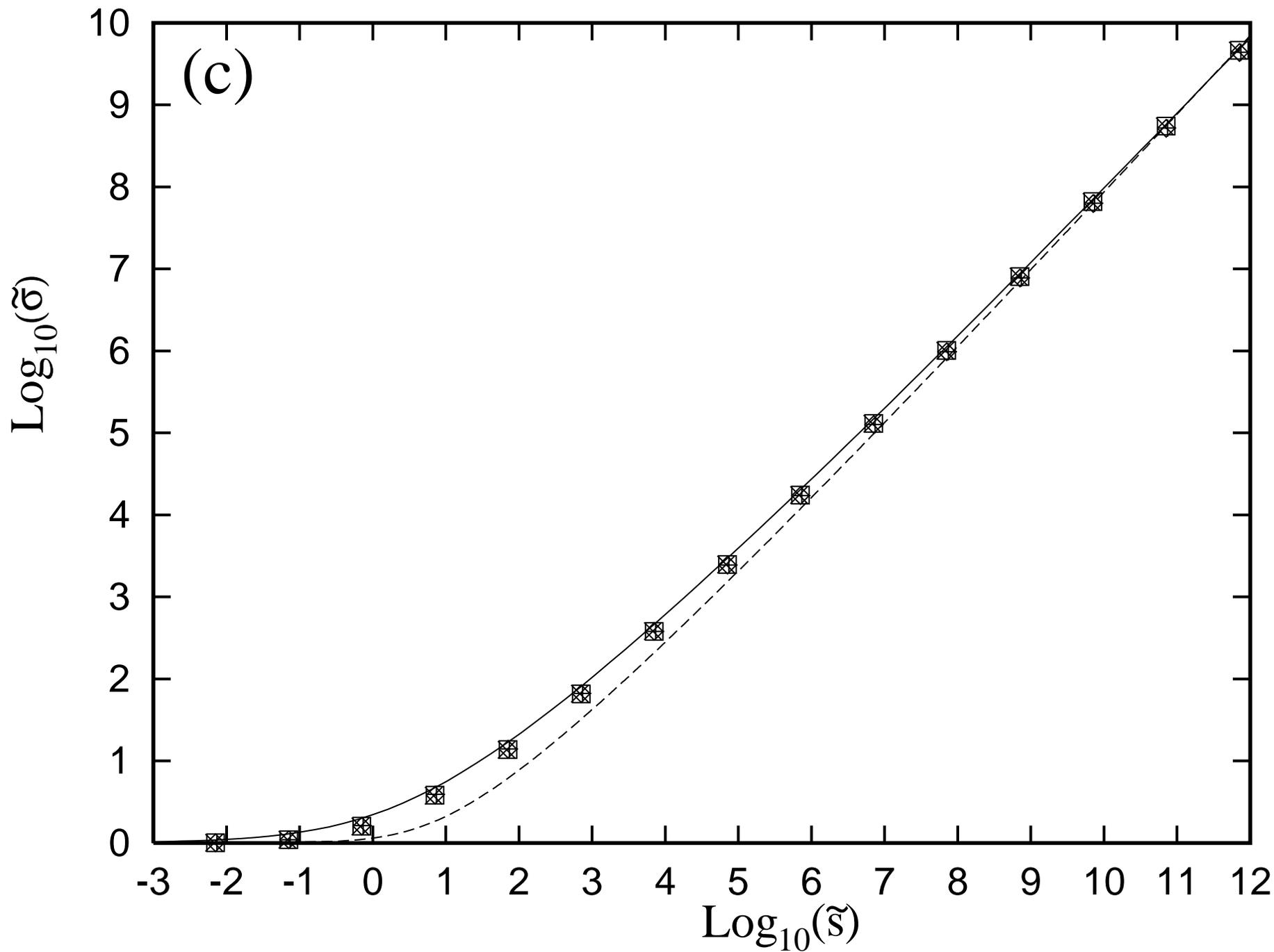
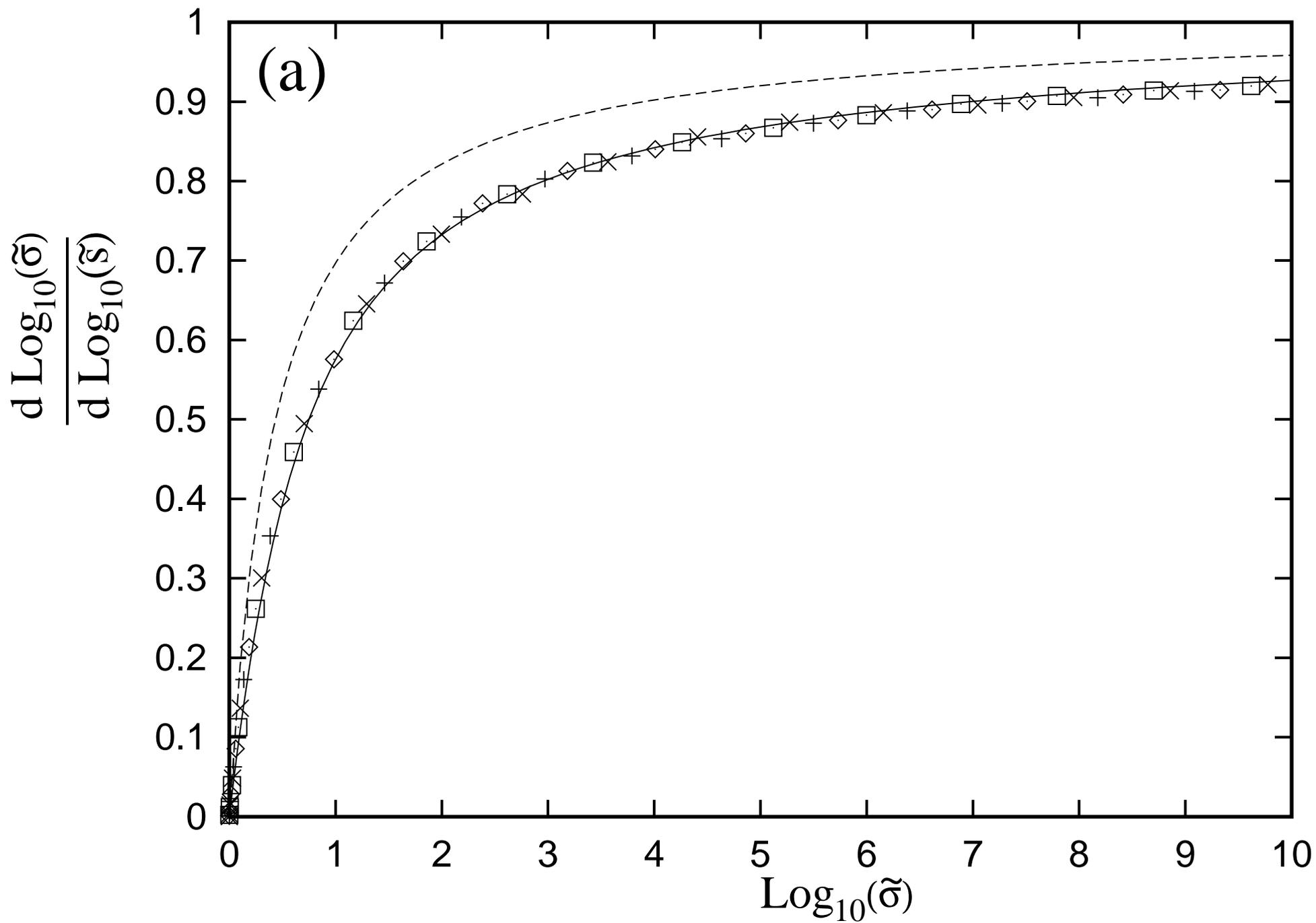
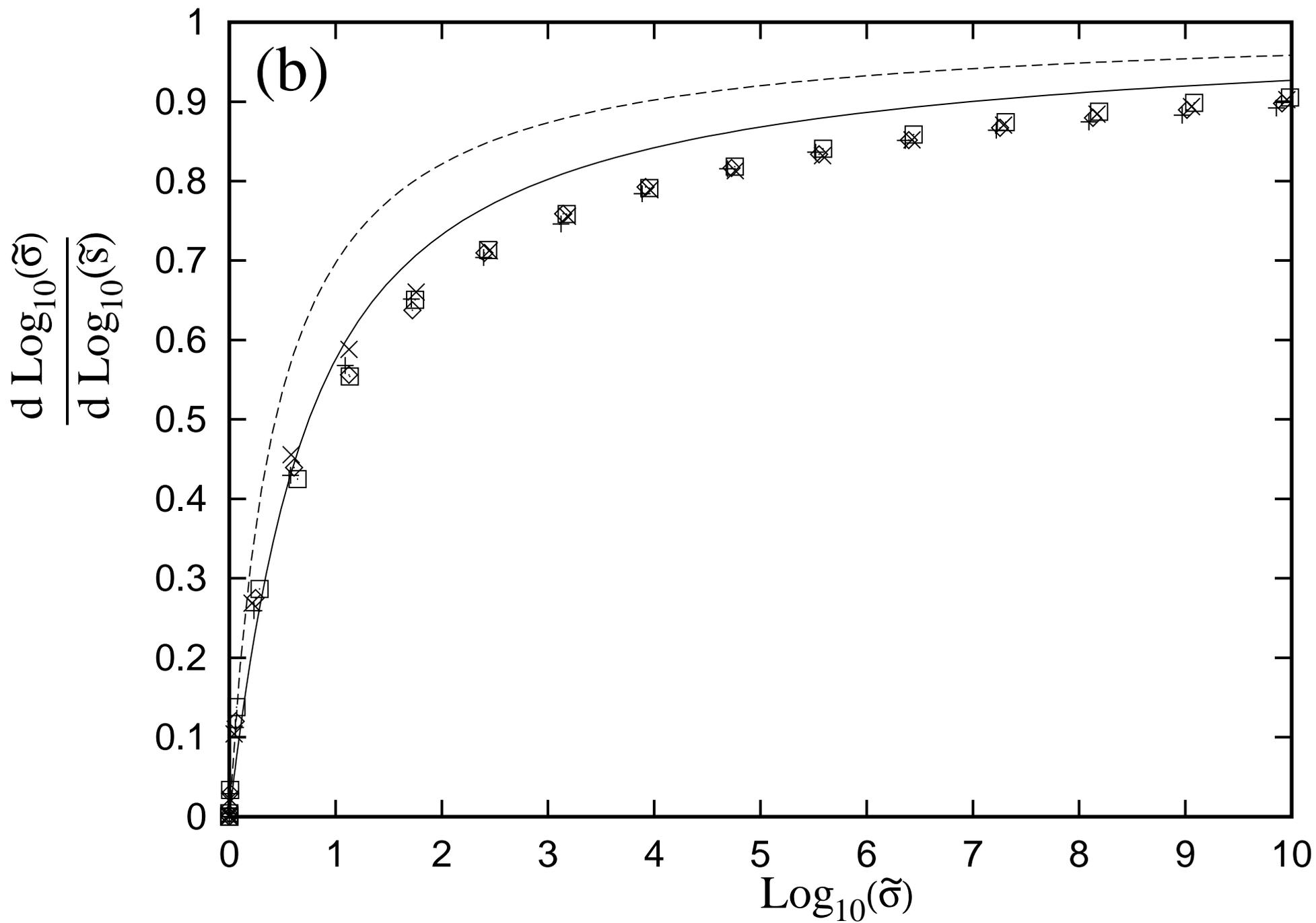
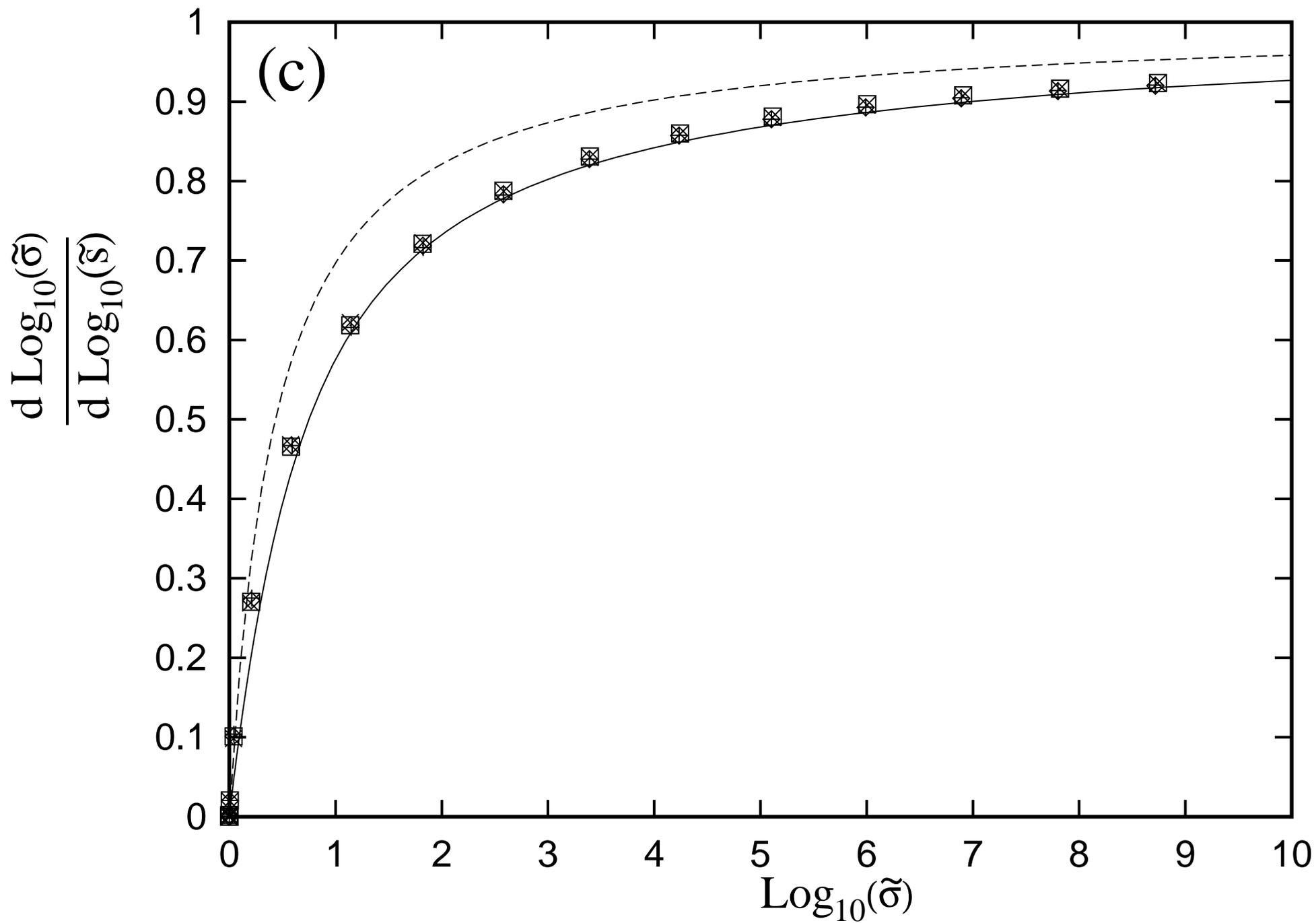
\end{document}